\RequirePackage{fix-cm}
\documentclass[twocolumn]{svjour3}          

\smartqed  

\usepackage{graphicx}
\usepackage{latexsym}
\usepackage{amsmath}
\usepackage{amssymb}
\usepackage{multirow}
\usepackage{cuted}

\journalname{Brazilian Journal of Physics}

\begin{document}

\title{Non-monotonicity of trace distance under tensor products}
\author{Jonas Maziero }

\institute{Jonas Maziero \at Departamento de F\'isica, Universidade Federal de Santa Maria, 97105-900, Santa Maria, RS, Brazil \\ \email{jonas.maziero@ufsm.br}}

\date{Received: date / Accepted: date}

\maketitle

\begin{abstract}
The trace distance (TD) possesses several of the good properties required
for a faithful distance measure in the quantum state space. Despite
its importance and ubiquitous use in quantum information science,
one of its questionable features, its possible non-monotonicity under
taking tensor products of its arguments (NMuTP), has been hitherto
unexplored. In this article we advance analytical and numerical investigations
of this issue considering different classes of states living in a
discrete and finite dimensional Hilbert space. Our results reveal
that although this property of TD does not shows up for pure states
and for some particular classes of mixed states, it is present in
a non-negligible fraction of the regarded density operators. Hence,
even though the percentage of quartets of states leading to the NMuTP
drawback of TD and its strength decrease as the system's dimension
grows, this property of TD must be taken into account before using
it as a figure of merit for distinguishing mixed quantum states.

\keywords{Quantum distance measures \and Trace distance \and Monotonicity under tensor
products}

\end{abstract}


\maketitle

\section{Introduction}

Quantifiers for the distance (distinguishability) between two density
operators in the quantum state space $\mathcal{D}(\mathcal{H})$--the
space formed by positive semidefinite matrices with trace equal to
one--are an essential and frequently used item in the quantum information
scientist toolkit \cite{Nielsen_Book,Wilde_Book,Fuchs_PhD}. For instance,
how well a quantum system was prepared \cite{Watanabe_QSP}, manipulated
\cite{Walter_EOWQC} or protected \cite{Laflamme_QEC} in an experiment
is usually evaluated via how close (how indistinguishable) its real
state is from what one would ideally expect. Distance measures in
$\mathcal{D}(\mathcal{H})$ naturally appear also in the contexts
of quantum foundations \cite{Gisin_cloning,Wilde_U_Mem,Zurek_qcb1},
quantum processes \cite{Cory_QP,Nielsen_QP}, quantum cryptography
\cite{Fuchs_Cryp,Gisin_RevQC}, quantum phase transitions \cite{Gu_RevF},
quantum speed limits \cite{Plastino_QSL,Kok_QSL,Davidovich_QSL,Lutz_QSL,Plenio_QSL,Fan_QSL},
quantum channel capacities \cite{Giovannetti_Channels}, and also
in the theories of quantum entanglement \cite{Horodecki_RevE,Davidovich_RevE},
quantum discord \cite{Lucas_RevD,Vedral_RevD,Streltsov_B}, and quantum
coherence \cite{Plenio_QC,Aberg_QC,Girolami_QC,Zhao_QC}.

Several distance (distinguishability) measures in the quantum state space
have been proposed in the literature in the last decades. A partial
list is provided in Ref. \cite{Audenaerte_DM}. A few examples are
the Bures' distance, that is defined in terms of a similarity measure
known as Uhlmann's fidelity \cite{Bures,Uhlmann_F} (for a critical
assessment regarding the use of this function in quantum information
science see Ref. \cite{Paris_F}), the $p-$norm distance, with the
trace distance (or $1-$norm distance) and the Hilbert-Schmidt distance
(or $2-$norm distance) being used more frequently (see Refs. \cite{Petz_p_norm,Sarandy_1Norm_D,Sarandy_DSC_exp,Vedral_2Norm_D}
and references therein), the quantum relative entropy \cite{Umegaki_QRE,Petz_QRE,Vedral_QRE,Maziero_dist_MI},
and the quantum Chernoff bound \cite{Szkola_QCB,Acin_QCB}, with these
last two distinguishability measures being defined operationally,
respectively, in the contexts of asymmetric and symmetric quantum
hypothesis testing.

In this article we are interested mainly in one of the most popular
distance measures, the trace distance, that is defined using the trace
norm. For a Hermitian matrix $A$, the \emph{trace norm} is defined
and given as follows:
\begin{equation}
||A||_{1}:=\mathrm{Tr}\sqrt{A^{\dagger}A}=\mathrm{Tr}\sqrt{A^{2}}={\textstyle \sum_{j}}|a_{j}|,\label{eq:trace_norm}
\end{equation}
with $|a_{j}|$ being the absolute value of the real eigenvalues of
$A$. We can quantify how dissimilar two density operators $\rho$
and $\zeta$ are by their \emph{trace distance} (TD), which is defined
as the trace norm of their subtraction,
\begin{equation}
d_{tr}(\rho,\zeta):=||\rho-\zeta||_{1},\label{eq:trace_distance}
\end{equation}
and assumes values between zero and two \cite{Wilde_Book}. 

This mathematical function possesses several of those properties required
for a faithful distance (distinguishability) measure in the quantum
state space \cite{Nielsen_Book,Wilde_Book}: For the density operators
$\rho$, $\zeta$, and $\alpha$, the trace distance is, e.g., positive
semidefinite ($d_{tr}(\rho,\zeta)\ge0$), it is zero if and only if
the two density operators are equal ($d_{tr}(\rho,\zeta)=0\Leftrightarrow\rho=\zeta$),
it is symmetric ($d_{tr}(\rho,\zeta)=d_{tr}(\zeta,\rho)$), it obeys
the triangle inequality ($d_{tr}(\rho,\zeta)\le d_{tr}(\rho,\alpha)+d_{tr}(\alpha,\zeta)$),
it is invariant under unitary transformations ($d_{tr}(\rho,\zeta)=d_{tr}(U\rho U^{\dagger},U\zeta U^{\dagger})$
for $UU^{\dagger}=\mathbb{I}_{d}$, where $\mathbb{I}_{d}$ is the
$d\mathrm{x}d$ identity matrix), it leads to the equality $d_{tr}(\rho\otimes\alpha,\zeta\otimes\alpha)=d_{tr}(\rho,\zeta)$,
it is monotonic under discarding subsystems ($d_{tr}(\rho_{1},\zeta_{1})\le d_{tr}(\rho_{12},\zeta_{12})$
with $x_{1}=\mathrm{Tr}_{2}(x_{12})$), and it is consequently also
monotonic under trace-preserving quantum operations ($d_{tr}(\rho,\zeta)\ge d_{tr}(\Phi(\rho),\Phi(\zeta))$
with $\Phi(x)={\textstyle \sum_{j}}K_{j}xK_{j}^{\dagger}$ and ${\textstyle \sum_{j}}K_{j}^{\dagger}K_{j}=\mathbb{I}_{d}$).

Notwithstanding, it was mentioned in Ref. \cite{Acin_QCB} that the
trace distance lacks monotonicity under taken tensor products of its
arguments. That is to say, we can find four density operators $\rho$,
$\zeta$, $\xi$, and $\eta$ such that the following inequalities
are satisfied:
\begin{eqnarray}
d_{tr}(\rho,\zeta) & \gtrless & d_{tr}(\xi,\eta),\label{eq:ineq0}\\
 & \mbox{and}\nonumber \\
d_{tr}(\rho^{\otimes2},\zeta^{\otimes2}) & \lessgtr & d_{tr}(\xi^{\otimes2},\eta^{\otimes2}).\label{eq:ineq1}
\end{eqnarray}
This non-monotonicity under tensor products (NMuTP) does not seem
to be a desirable property for a distance measure in $\mathcal{D}(\mathcal{H})$.
If a pair of states of a quantum system is more distinguishable than
another pair of states, one would expect the same to hold for two
identical and uncorrelated copies of the system prepared in those
states. 

Two relevant questions to answer regarding this issue are (i) for
what kind of state and (ii) how often the inequalities in Eqs. (\ref{eq:ineq0})
and (\ref{eq:ineq1}) can simultaneously hold. The remainder of this
article will be devoted to answer these questions for the cases of
general state vectors (Sec. \ref{pure}), for one-qubit states (Sec.
\ref{mixed_1qb}), and also for high-dimensional quantum systems (Sec.
\ref{mixed_1qd}).

\section{The non-monotonicity of trace distance under tensor products}

This section is dedicated to investigate such an issue considering
some particular classes of states. Though we present some analytical
results, much of the work should be numeric. We will start using general
pure states and a two-level quantum system to address the question
(i). In the sequence the question (ii) will be studied mainly with
regard to its dependence with the system's dimension.

\subsection{Arbitrary pure states}

\label{pure}

Let $\rho$, $\zeta$, $\xi$, and $\eta$ be arbitrary state vectors
on the discrete Hilbert space $\mathcal{H}$ of dimension $d$. The
trace distance between the pair of states $x=\rho\mbox{ }(\xi)$ and
$y=\zeta\mbox{ }(\eta)$ can be written as \cite{Wilde_Book}:
\begin{equation}
d_{tr}(x,y)=2\sqrt{1-\mathrm{Tr}(xy)}.
\end{equation}
Given that $0\le\mathrm{Tr}(xy)\le1$, $d_{tr}(\rho,\zeta)>d_{tr}(\xi,\eta)$
implies $\mathrm{Tr}(\rho\zeta)<\mathrm{Tr}(\xi\eta)$. Using this
inequality and the fact that, in the present case,
\begin{equation}
d_{tr}(x^{\otimes2},y^{\otimes2})=2\sqrt{1-[\mathrm{Tr}(xy)]^{2}},
\end{equation}
we see that $d_{tr}(\rho^{\otimes2},\zeta^{\otimes2})>d_{tr}(\xi^{\otimes2},\eta^{\otimes2}).$
Thus, if all the states involved are pure states, the trace distance
does not suffer from the NMuTP drawback under analysis here.

\subsection{One-qubit states}

\label{mixed_1qb}

\subsubsection{Collinear States}

\label{mixed_1qb_1}

Let us consider the special case in which the pairs of density operators
$(\rho,\zeta)$ and $(\xi,\eta)$ are, individually, collinear. That
is to say, let e.g. 
\begin{equation}
\rho=2^{-1}(\mathbb{I}_{2}+\vec{r}\cdot\vec{\sigma})\mbox{ and }\zeta=2^{-1}(\mathbb{I}_{2}+\vec{z}\cdot\vec{\sigma})
\end{equation}
with the two Bloch's vectors being $\vec{r}=r\hat{n}$ and $\vec{z}=\pm z\hat{n}$,
where $\hat{n}$ is any unit vector in $\mathbb{R}^{3}$ and $\vec{\sigma}$
is the Pauli's vector. One can readily show that
\begin{equation}
d_{tr}(\rho,\zeta)=|r\mp z|.
\end{equation}

For the tensor products we have
\begin{eqnarray}
\rho^{\otimes2}-\zeta^{\otimes2} & = & 2^{-2}((r\mp z)(\mathbb{I}_{2}\otimes\hat{n}\cdot\vec{\sigma}+\hat{n}\cdot\vec{\sigma}\otimes\mathbb{I}_{2})\nonumber \\
 &  & +(r^{2}-z^{2})\hat{n}\cdot\vec{\sigma}\otimes\hat{n}\cdot\vec{\sigma})\\
 & = & 2^{-2}((r\mp z)(r\pm z+2)P_{+}\otimes P_{+}\nonumber \\
 &  & +(r\mp z)(r\pm z-2)P_{-}\otimes P_{-}\nonumber \\
 &  & -(r^{2}-z^{2})(P_{+}\otimes P_{-}+P_{-}\otimes P_{+})),\label{eq:rrzz}
\end{eqnarray}
where we used $\hat{n}\cdot\vec{\sigma}=P_{+}-P_{-}$ and $\mathbb{I}_{2}=P_{+}+P_{-}$.
It is straightforward applying Eq. (\ref{eq:rrzz}) to get
\begin{eqnarray}
d_{tr}(\rho^{\otimes2},\zeta^{\otimes2}) & = & d_{tr}(\rho,\zeta)2^{-1}(2+|r\pm z|).
\end{eqnarray}
We see that $d_{tr}(\rho^{\otimes2},\zeta^{\otimes2})$ is a monotonically
increasing function of $d_{tr}(\rho,\zeta)$. Thus, for this particular
set of states, the inequalities in Eqs. (\ref{eq:ineq0}) and (\ref{eq:ineq1})
cannot be satisfied simultaneously.

\subsubsection{$\rho$, $\zeta$, $\xi$, and $\eta$ arbitrary states}

\begin{figure}
\begin{centering}
\includegraphics[scale=0.4]{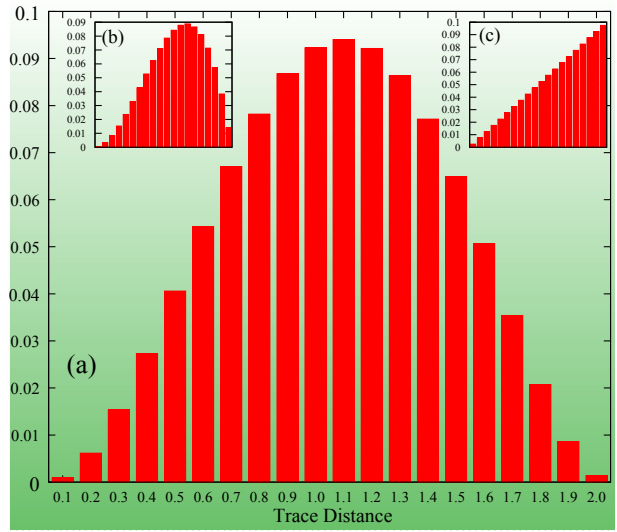}
\par\end{centering}

\protect\caption{(color online) Probability distribution for the different values of
trace distance ($d_{tr}(\rho,\zeta)=||\vec{r}-\vec{z}||_{2}$, where
$\zeta=2^{-1}(\mathbb{I}_{2}+\vec{z}\cdot\vec{\sigma})$) for $10^{8}$
pairs of randomly-generated one-qubit states: (a) two density operators
$(\rho\mbox{, }\zeta)$, (b) one mixed and one pure state $(\rho\mbox{, }|\zeta\rangle)$,
and (c) two state vectors $(|\rho\rangle\mbox{, }|\zeta\rangle)$.
The mean value of TD in the three cases are, respectively, $1.03$,
$1.20$, and $1.33$.}

\label{prob_dist}
\end{figure}

Even this apparently simple one-qubit case is not easily tamable for
analytical computations. Hence we recourse to numerical calculations
via Monte Carlo (random) sampling of the quartets of states to be
used. The computations of eigenvalues involved in this article are
done utilizing the LAPACK subroutines (see Ref. \cite{LAPACK}). Let
us start by using the Fano's parametrization \cite{Fano_1983} to
write an one-qubit density matrix $x=\rho\mbox{, }\zeta\mbox{, }\xi\mbox{, }\eta$
in the form:
\begin{equation}
x=2^{-1}(\mathbb{I}_{2}+{\textstyle \sum_{j=1}^{3}}x_{j}\sigma_{j}),\label{eq:1qbDM}
\end{equation}
with $\vec{x}=(x_{1},x_{2},x_{3})$, where $x_{1}=||\vec{x}||_{2}\sin\theta\cos\phi$,
$x_{2}=||\vec{x}||_{2}\sin\theta\sin\phi$, and $x_{3}=||\vec{x}||_{2}\cos\theta$.
The parameters appearing in these equations can assume values in the
ranges \cite{Fano_1983,Petruccione_PofU}: $||\vec{x}||_{2}\in[0,1]$,
$\theta\in[0,\pi]$, and $\phi\in[0,2\pi]$. In order to obtain an
uniform distribution of points (states) in the Bloch's ball, each
one of the quantum states is generated setting 
\begin{equation}
||\vec{x}||_{2}=(t_{1})^{1/3}\mbox{, }\theta=\arccos(-1+2t_{2})\mbox{, }\phi=2\pi t_{3}
\end{equation}
with $t_{j}$ ($j=1,2,3$) being a pseudorandom number with uniform
distribution in the interval $[0,1]$. The Mersenne Twister pseudo-random
number generator \cite{Matsumoto_MT} is applied to produce these
numbers. By setting the Euclidean norm of the Bloch's vector equal
to one (zero) we obtain pure (maximally mixed) states. For the sake
of illustration, the probability distribution for the values of TD
between pairs of randomly-generated one-qubit states is presented
in Fig. \ref{prob_dist}. 

It is worthwhile mentioning at this point that we have made several
tests from which we found that the numerical and analytical results
for the TD coincide up to the fifteenth digit when applied to random
states in those classes considered in Secs. \ref{mixed_1qb_1} and
\ref{pure}. More specifically, we generated one million pairs of random collinear states ($d=2$) and one million pairs of random pure states (see e.g. Ref. \cite{Maziero_pRPV}) for each value of the system dimension (with $d = 2,\cdots,20$). The error, in each case, is computed by comparing the trace distance obtained via diagonalization with the LAPACK subroutines and the value of TD obtained using its analytical expression. Then the precision is established via the worst case error.

\begin{table}
\begin{centering}
\begin{tabular}{|c|c|c|c|c|}
\hline 
States generated & Percentage & $\langle G\rangle$ & $\Delta G$ & $G_{\mathrm{max}}$\tabularnewline
\hline 
\hline 
$(\rho,\zeta),(\xi,\eta)$ & $7.28$ & $0.161$ & $0.083$ & $0.475$\tabularnewline
\hline 
$(\rho,\zeta),(\xi,|\eta\rangle)$ & $7.86$ & $0.186$ & $0.091$ & $0.486$\tabularnewline
\hline 
$(\rho,|\zeta\rangle),(\xi,|\eta\rangle)$ & $3.16$ & $0.071$ & $0.038$ & $0.190$\tabularnewline
\hline 
$(\rho,\zeta),(|\xi\rangle,|\eta\rangle)$ & $4.06$ & $0.134$ & $0.072$ & $0.333$\tabularnewline
\hline 
$(\rho,|\zeta\rangle),(|\xi\rangle,|\eta\rangle)$ & $3.00$ & $0.083$ & $0.044$ & $0.194$\tabularnewline
\hline 
$(\rho,\zeta),(\xi,\mathbb{I}_{2}/2)$ & $8.49$ & $0.192$ & $0.098$ & $0.488$\tabularnewline
\hline 
$(\rho,\zeta),(|\xi\rangle,\mathbb{I}_{2}/2)$ & $20.75$ & $0.226$ & $0.096$ & $0.500$\tabularnewline
\hline 
$(\rho,|\zeta\rangle),(\xi,\mathbb{I}_{2}/2)$ & $3.20$ & $0.101$ & $0.045$ & $0.248$\tabularnewline
\hline 
$(\rho,|\zeta\rangle),(|\xi\rangle,\mathbb{I}_{2}/2)$ & $7.67$ & $0.123$ & $0.037$ & $0.177$\tabularnewline
\hline 
$(|\rho\rangle,|\zeta\rangle),(\xi,\mathbb{I}_{2}/2)$ & $2.63$ & $0.107$ & $0.043$ & $0.177$\tabularnewline
\hline 
$(|\rho\rangle,|\zeta\rangle),(|\xi\rangle,\mathbb{I}_{2}/2)$ & $8.85$ & $0.169$ & $0.004$ & $0.177$\tabularnewline
\hline 
\end{tabular}
\par\end{centering}

\protect\caption{Percentage of the $10^{6}$ quartets of randomly-generated one-qubit
states leading to the NMuTP drawback of trace distance. In the
last three columns are presented the average value, standard deviation,
and maximum value of the strength of the NMuTP drawback of TD, as
defined in Eq. (\ref{eq:gap}), for each case study.}

\label{table_1qb}
\end{table}

\begin{figure}
\begin{centering}
\includegraphics[scale=0.35]{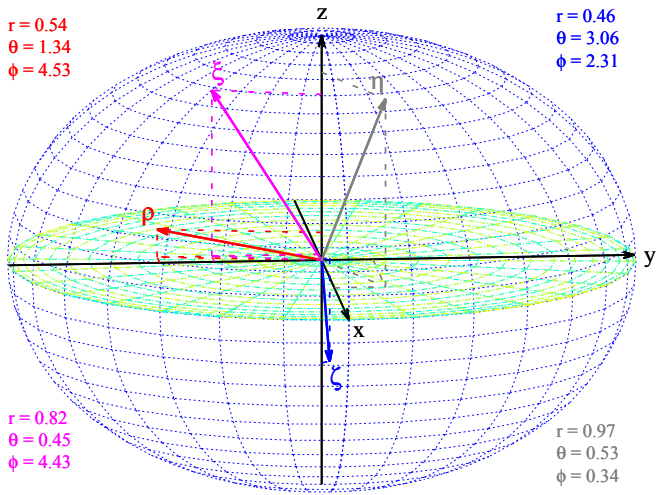}
\par\end{centering}

\protect\caption{(color online) Example of a quartet of states for which the trace
distance is not monotonic under taking tensor products of its arguments.
Here $r$ is the size of the corresponding Bloch's vector, the angles
are given in radians, and $d_{tr}(\rho,\zeta)=0.80$, $d_{tr}(\xi,\eta)=0.76$,
$d_{tr}(\rho^{\otimes2},\zeta^{\otimes2})=0.87$, and $d_{tr}(\xi^{\otimes2},\eta^{\otimes2})=1.07$.}

\label{exe_BlochS}
\end{figure}

We proved in the previous subsections that if the pairs $(\rho,\zeta)$
and $(\xi,\eta)$ are, individually, collinear or if all the four
states are pure, we shall have no NMuTP drawback of trace distance.
However, as is shown in Table \ref{table_1qb}, for all the other
possibilities a significant fraction of the one million one-qubit
quartets of states randomly generated presented this unwanted property
of TD. In Fig. \ref{exe_BlochS} we draw an example of such a quartet
of states.

\begin{figure}
\begin{centering}
\includegraphics[scale=0.52]{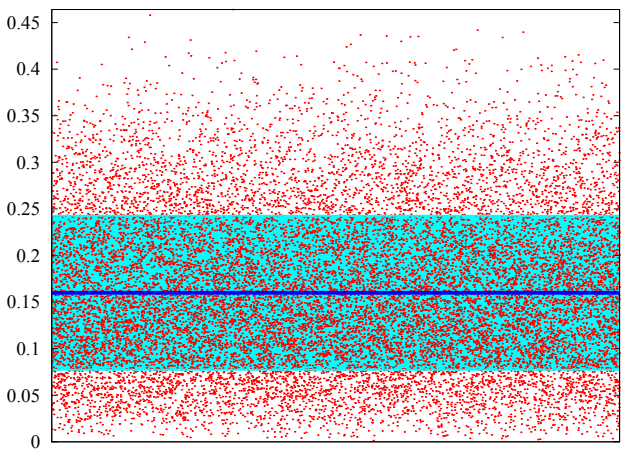}
\par\end{centering}

\protect\caption{(color online) Values of the strength defined in Eq. (\ref{eq:gap})
for one fourth of the $72802$ quartets of states leading to the NMuTP
drawback of TD in the case study $(\rho,\zeta),(\xi,\eta)$, in one
numerical experiment with one million random samples generated. The
blue line is $\langle G\rangle$ and the points in the cyan area are
values of $G$ in the interval $[\langle G\rangle-\Delta G,\langle G\rangle+\Delta G]$
(see Table \ref{table_1qb}).}

\label{fig_distG}
\end{figure}

For the sake of measuring the \emph{strength of the NMuTP drawback
of TD}, when applicable, we will define the following quantity:
\begin{eqnarray}
G(\rho,\zeta,\xi,\eta) & := & |d_{tr}(\rho,\zeta)-d_{tr}(\xi,\eta)|\label{eq:gap}\\
 &  & +|d_{tr}(\rho^{\otimes2},\zeta^{\otimes2})-d_{tr}(\xi^{\otimes2},\eta^{\otimes2})|.\nonumber 
\end{eqnarray}
The quantity $G$ measures how far the TD is from been monotonic under tensor products. As $G$ is defined only for those quartets of states leading to the NMuTP of TD, its lower bound is zero.
In order to access more details about the distribution of $G$, we
shall use its mean value $\langle G\rangle$, standard deviation $\Delta G$,
and maximum value $G_{\mathrm{max}}$. These quantities are also shown
in Table \ref{table_1qb}. A sample of the values of $G$ for the
case study $(\rho,\zeta),(\xi,\eta)$ is presented in Fig. \ref{fig_distG}.

Even with these additional informations, as can be seem in Table \ref{table_1qb},
in the general case the relationship between the existence of the
NMuTP drawback of TD and the classes of states involved is not an
easy matter. For instance, starting with general states and then restricting
one of them to be pure we pass from a percentage of $7.28\mbox{ \%}$
to $7.86\mbox{ \%}$. But then the addition of the same restriction
for one state of the other pair reduces the percentage with the undesired
property of TD to $3.16\mbox{ \%}$. Several other similar nontrivial
changes in the percentages can be identified. One striking one is
that in the last line of the table. For four pure states there is
no drawback, however just by putting one of the states in the center
of the Bloch's ball, we get a percentage of $8.85\mbox{ \%}$, the
second higher among the classes of one-qubit states studied. These
results stress the richness and complexity of the quantum state space,
already for the composition of two of its simplest systems. Thus,
in order to simplify the analysis, we will investigate in the next
section the general dependence of the NMuTP drawback of TD with the
dimension of the system.

\subsection{General one-qudit states}

\label{mixed_1qd}

\begin{figure}[b]
\begin{centering}
\includegraphics[scale=0.35]{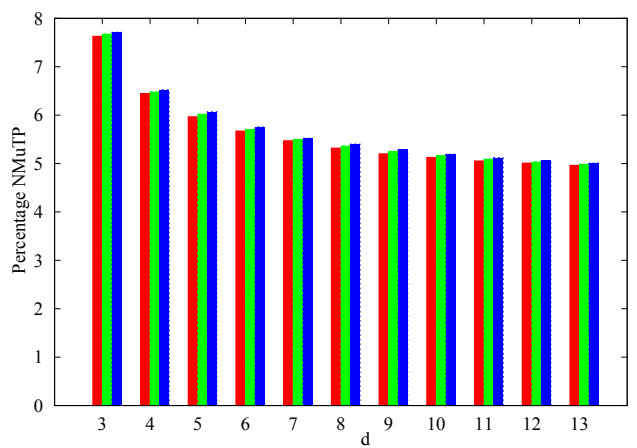}
\par\end{centering}

\protect\caption{(color online) Minimum value (red bar on the left), mean value (green
bar in the middle), and maximum value (blue bar on the right) of the
fraction of the quartets of randomly-generated quantum states leading
to the NMuTP drawback of trace distance as a function of the system's
dimension for ten numerical experiments. We generated one million
quartets of states in each experiment. }

\label{fig_nmutp}
\end{figure}

In this subsection we shall study the NMuTP drawback of trace distance
for $d-$level quantum systems, known as qudits. As there is no explicit
parametrization for density matrices with $d\ge3$ \cite{Petruccione_PofU},
we will proceed as follows. Let us first look for the spectral decomposition
of a given density operator $x=\rho\mbox{, }\zeta\mbox{, }\xi\mbox{, }\eta$:
\begin{equation}
x={\textstyle \sum_{j=1}^{d}x_{j}|x_{j}\rangle\langle x_{j}|}.
\end{equation}
Once the eigenvalues of $x$ form a probability distribution, i.e.,
\begin{equation}
x_{j}\ge0\mbox{ and }{\textstyle \sum_{j=1}^{d}x_{j}=1,}
\end{equation}
we can use a geometric parametrization for them \cite{Fritzche4}:
\begin{equation}
x_{j}=\sin^{2}\theta_{j-1}{\textstyle \prod_{k=j}^{d-1}}\cos^{2}\theta_{k}
\end{equation}
with $\theta_{0}=\pi/2$. The details about the numerical generation
of $\{x_{j}\}$ using this parametrization can be found in Ref. \cite{Maziero_pRPV}.

The basis formed by the eigenvectors of a density operator $x$, $\{|x_{j}\rangle\}$,
can be obtained from the computational basis, $\{|j\rangle\}$, using
an unitary matrix $U$, i.e.,
\begin{equation}
|x_{j}\rangle=U|j\rangle,\mbox{ with }j=1,\cdots,d.
\end{equation}
There are several parametrizations for unitary matrices \cite{Petruccione_PofU}.
Here we use the Hurwitz's parametrization with Euler's angles. For
details see e.g. Ref. \cite{Zyczkowski_U}. 

\begin{figure}[ht!]
\begin{centering}
\includegraphics[scale=0.85]{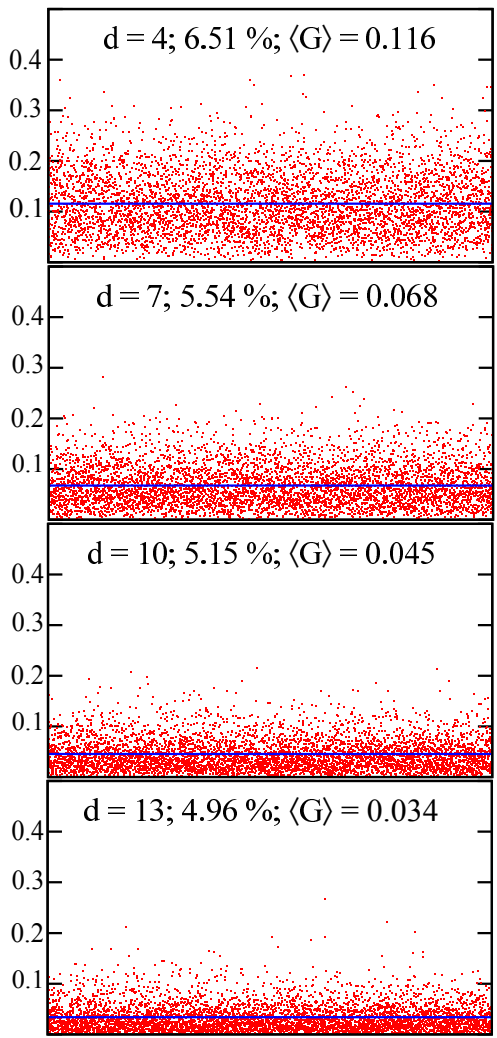}
\par\end{centering}
\protect\caption{(color online) Samples with 5000 values of the strength of the NMuTP
drawback of trace distance for some values of the system's dimension
$d$. The NMuTP average percentage of the whole sample with ten sets
of $10^{6}$quartets of states is shown at the side of $d$. The blue
line indicates $\langle G\rangle$.}
\label{fig_distG_d}
\end{figure}

By creating pseudo-random probability distributions $\{x_{j}\}$ and
pseudo-random unitary matrices $U$, we did ten numerical experiments
for each value of the system's dimension $d$ generating one million
quartets of states in each experiment. The mean, minimum, and maximum
values of the percentages of the quartets of states leading to NMuTP
of TD are shown in Fig. \ref{fig_nmutp}. In the Fig. \ref{fig_distG_d}
we present samples of the distribution of values of the drawback's
strength for some values of $d$. We can see in these figures a steady
decreasing of such a proportion and strength as the system dimension
$d$ grows.

\section{Final remarks}

The trace distance has several good properties that rank it as one
of the major distance measures between quantum states. Nevertheless,
it also presents a potential drawback, the possibility of being non-monotonic
under taking tensor products of its arguments, that was shown in this
article to exist for a non-negligible fraction of the density matrices
investigated. Thus, although such issue seems not to be much relevant
for high-dimensional quantum systems, it must be taken into account
when dealing with few qubits. 

The important question that yet remains is if, in the cases were the NMuTP of TD is significant, it has some undesirable consequence for important functions in quantum information science.
The possible implications of this issue regarding, for instance, the
quantification of quantum entanglement, of quantum discord, and of
quantum coherence is an appealing topic for further researches. It
would also be fruitful analyzing the NMuTP drawback considering other
quantum distance measures. The obtention of a more precise operational and/or physical interpretation of $G$ and its upper bound are also left as open problems.

\begin{acknowledgements}
This work was supported by the Brazilian funding agencies: National Counsel of Technological and Scientific Development (CNPq), under process number 303496$\slash$2014-2 and under grant number 441875/2014-9, and by the National Institute of Science and Technology - Quantum Information (INCT-IQ), under process number 2008/57856-6.
\end{acknowledgements}


\begin{thebibliography}{10}
\bibitem{Nielsen_Book} M.A. Nielsen and I.L. Chuang, \emph{Quantum
Computation and Quantum Information} (Cambridge University Press,
Cambridge, 2000)

\bibitem{Wilde_Book} M.M. Wilde, \emph{Quantum Information Theory}
(Cambridge University Press, Cambridge, 2013)

\bibitem{Fuchs_PhD} C.A. Fuchs, Distinguishability and accessible
information in quantum theory, arXiv:quant-ph/9601020

\bibitem{Watanabe_QSP} P. Neumann, N. Mizuochi, F. Rempp, P. Hemmer,
H. Watanabe, S. Yamasaki, V. Jacques, T. Gaebel, F. Jelezko, J. Wrachtrup,
Multipartite entanglement among single spins in diamond, Science \textbf{320},
1326 (2008)

\bibitem{Walter_EOWQC} P. Walther, K.J. Resch, T. Rudolph, E. Schenck,
H. Weinfurter, V. Vedral, M. Aspelmeyer, and A. Zeilinger, Experimental
one-way quantum computing, Nature \textbf{434}, 169 (2005)

\bibitem{Laflamme_QEC} J. Zhang, R. Laflamme, and D. Suter, Experimental
implementation of encoded logical qubit operations in a perfect quantum
error correcting code, Phys. Rev. Lett. \textbf{109}, 100503 (2012)

\bibitem{Gisin_cloning} V. Scarani, S. Iblisdir, N. Gisin, and A. Ac\'in,
Quantum cloning, Rev. Mod. Phys. \textbf{77}, 1225 (2005)

\bibitem{Wilde_U_Mem} A.K. Pati, M.M. Wilde, A.R.U. Devi, A.K. Rajagopal, and Sudha, Quantum discord and classical correlation
can tighten the uncertainty principle in the presence of quantum memor,
Phys. Rev. A \textbf{86}, 042105 (2012)

\bibitem{Zurek_qcb1} M. Zwolak, C.J. Riede, and W.H. Zurek, Amplification,
redundancy, and quantum chernoff information, Phys. Rev. Lett. \textbf{112},
140406 (2014)

\bibitem{Cory_QP} N. Boulant, T.F. Havel, M.A. Pravia, and D.G.
Cory, Robust method for estimating the Lindblad operators of a dissipative
quantum process from measurements of the density operator at multiple
time points, Phys. Rev. A \textbf{67}, 042322 (2003)

\bibitem{Nielsen_QP} A. Gilchrist, N.K. Langford, and M.A. Nielsen,
Distance measures to compare real and ideal quantum processes, Phys.
Rev. A \textbf{71}, 062310 (2005)

\bibitem{Fuchs_Cryp} C.A. Fuchs and J. van de Graaf, Cryptographic
distinguishability measures for quantum-mechanical states, IEEE Trans.
Inf. Theory \textbf{45}, 1216 (1999)

\bibitem{Gisin_RevQC} N. Gisin, G. Ribordy, W. Tittel, and H. Zbinden,
Quantum cryptography, Rev. Mod. Phys. \textbf{74}, 145 (2002)

\bibitem{Gu_RevF} S.-J. Gu, Fidelity approach to quantum phase transitions,
Int. J. Mod. Phys. B \textbf{24}, 4371 (2010)

\bibitem{Plastino_QSL} A. Borr\'as, M. Casas, A.R. Plastino, and
A. Plastino, Entanglement and the lower bounds on the speed of quantum
evolution, Phys. Rev. A \textbf{74}, 022326 (2006)

\bibitem{Kok_QSL} P.J. Jones and P. Kok, Geometric derivation of
the quantum speed limit, Phys. Rev. A \textbf{82}, 022107 (2010)

\bibitem{Davidovich_QSL} M.M. Taddei, B.M. Escher, L. Davidovich,
and R.L. de Matos Filho, Quantum speed limit for physical processes,
Phys. Rev. Lett. \textbf{110}, 050402 (2013)

\bibitem{Lutz_QSL} S. Deffner and E. Lutz, Quantum speed limit for
non-Markovian dynamics, Phys. Rev. Lett. \textbf{111}, 010402 (2013)

\bibitem{Plenio_QSL} A. del Campo, I.L. Egusquiza, M.B. Plenio,
and S.F. Huelga, Quantum speed limits in open system dynamics, Phys.
Rev. Lett. \textbf{110}, 050403 (2013)

\bibitem{Fan_QSL} Y.-J. Zhang, W. Han, Y.-J. Xia, J.-P. Cao, and
H. Fan, Quantum speed limit for arbitrary initial states, Sci. Rep.
\textbf{4}, 4890 (2014)

\bibitem{Giovannetti_Channels} F. Caruso, V. Giovannetti, C. Lupo,
and S. Mancini, Quantum channels and memory effects, Rev. Mod. Phys.
\textbf{86}, 1203 (2014)

\bibitem{Horodecki_RevE} R. Horodecki, P. Horodecki, M. Horodecki,
and K. Horodecki, Quantum entanglement, Rev. Mod. Phys. \textbf{81},
865 (2009)

\bibitem{Davidovich_RevE} L. Aolita, F. de Melo, and L. Davidovich,
Open-system dynamics of entanglement: a key issues review, Rep. Prog.
Phys. \textbf{78}, 042001 (2015)

\bibitem{Lucas_RevD} L.C. C\'eleri, J. Maziero, and R.M. Serra,
Theoretical and experimental aspects of quantum discord and related
measures, Int. J. Quant. Inf. \textbf{9}, 1837 (2011)

\bibitem{Vedral_RevD} K. Modi, A. Brodutch, H. Cable, T. Paterek,
and V. Vedral, The classical-quantum boundary for correlations: Discord
and related measures, Rev. Mod. Phys. \textbf{84}, 1655 (2012)

\bibitem{Streltsov_B} A. Streltsov, \emph{Quantum Correlations Beyond
Entanglement} \emph{and Their Role in Quantum Information Theory}
(Springer, Berlin, 2015)

\bibitem{Plenio_QC} T. Baumgratz, M. Cramer, and M.B. Plenio, Quantifying
coherence, Phys. Rev. Lett. \textbf{113}, 140401 (2014)

\bibitem{Aberg_QC} J. \r{A}berg, Catalytic coherence, Phys. Rev.
Lett. \textbf{113}, 150402 (2014)

\bibitem{Girolami_QC} D. Girolami, Observable measure of quantum
coherence in finite dimensional systems, Phys. Rev. Lett. \textbf{113},
170401 (2014)

\bibitem{Zhao_QC} C.-s. Yu, Y. Zhang, and H. Zhao, Quantum correlation
via quantum coherence, Quant. Inf. Process. \textbf{13}, 1437 (2014)

\bibitem{Audenaerte_DM} K.M.R. Audenaert, Comparisons between quantum
state distinguishability measures, Quant. Inf. Comp. \textbf{14},
31 (2014)

\bibitem{Bures} D. Bures, An extension of Kakutani's theorem on infinite
product measures to the tensor product of semifinite $w^{*}-$algebras,
Trans. Amer. Math. Soc. \textbf{135}, 199 (1969)

\bibitem{Uhlmann_F} A. Uhlmann, The \textquotedblleft transition
probability\textquotedblright{} in the state space of a $^{\ast}-$algebra,
Rep. Math. Phys. \textbf{9}, 273 (1976)

\bibitem{Paris_F} M. Bina, A. Mandarino, S. Olivares, and M.G.A.
Paris, Drawbacks of the use of fidelity to assess quantum resources,
Phys. Rev. A \textbf{89}, 012305 (2014)

\bibitem{Petz_p_norm} D. P\'erez-Garc\'ia, M.M. Wolf, D. Petz,
and M.B. Ruskai, Contractivity of positive and trace-preserving maps
under Lp norms, J. Math. Phys. \textbf{47}, 083506 (2006)

\bibitem{Sarandy_1Norm_D} F.M. Paula, T.R. de Oliveira, and M.S. Sarandy, Geometric quantum discord through the Schatten 1-norm,
Phys. Rev. A \textbf{87}, 064101 (2013)

\bibitem{Sarandy_DSC_exp} F.M. Paula, I.A. Silva, J.D. Montealegre,
A.M. Souza, E.R. deAzevedo, R.S. Sarthour, A. Saguia, I.S. Oliveira,
D.O. Soares-Pinto, G. Adesso, and M.S. Sarandy, Observation of environment-induced
double sudden transitions in geometric quantum correlations, Phys.
Rev. Lett. \textbf{111}, 250401 (2013)

\bibitem{Vedral_2Norm_D} B. Daki\'c, V. Vedral, and \v{C}. Brukner,
Necessary and sufficient condition for nonzero quantum discord, Phys.
Rev. Lett. \textbf{105}, 190502 (2010)

\bibitem{Umegaki_QRE} H. Umegaki, Conditional expectation in an operator
algebra. IV. Entropy and information, Kodai Math. Sem. Rep. \textbf{14},
59 (1962)

\bibitem{Petz_QRE} F. Hiai and D. Petz, The proper formula for relative
entropy and its asymptotics in quantum probability, Comm. Math. Phys.
\textbf{143}, 99 (1991)

\bibitem{Vedral_QRE} V. Vedral, The role of relative entropy in quantum
information theory, Rev. Mod. Phys. \textbf{74}, 197 (2002)

\bibitem{Maziero_dist_MI} J. Maziero, Distribution of mutual information in multipartite states, Braz. J. Phys. \textbf{44}, 194 (2014)

\bibitem{Szkola_QCB} M. Nussbaum and A. Szko\l{}a, The Chernoff
lower bound for symmetric quantum hypothesis testing, Ann. Stat. \textbf{37},
1040 (2009)

\bibitem{Acin_QCB} K.M.R. Audenaert, J. Calsamiglia, R. Mun\~oz-Tapia,
E. Bagan, Ll. Masanes, A. Acin, and F. Verstraete, Discriminating
states: The quantum Chernoff bound, Phys. Rev. Lett. \textbf{98},
160501 (2007)

\bibitem{LAPACK} E. Anderson, Z. Bai, C. Bischof, S. Blackford, J.
Demmel, J. Dongarra, J. Du Croz, A. Greenbaum, S. Hammarling, A. McKenney,
and D. Sorensen, \emph{LAPACK Users' Guide,} 3rd Ed. (Society for
Industrial and Applied Mathematics, Philadelphia, 1999)

\bibitem{Fano_1983} U. Fano, Pairs of two-level systems, Rev. Mod.
Phys. \textbf{55}, 855 (1983)

\bibitem{Petruccione_PofU} E. Br\"{u}ning, H. M\"{a}kel\"{a}, A.
Messina, and F. Petruccione, Parametrizations of density matrices,
J. Mod. Opt. \textbf{59}, 1 (2012)

\bibitem{Matsumoto_MT} M. Matsumoto and T. Nishimura, Mersenne Twister:
A 623-dimensionally equidistributed uniform pseudorandom number generator,
ACM Trans. Model. Comput. Sim. \textbf{8}, 3 (1998)

\bibitem{Maziero_pRPV} J. Maziero, Generating pseudo-random discrete probability distributions,  Braz. J. Phys. \textbf{45}, 377 (2015)

\bibitem{Fritzche4} T. Radtke and S. Fritzsche, Simulation of n-qubit
quantum systems. IV. Parametrizations of quantum states, matrices
and probability distributions, Comput. Phys. Comm. \textbf{179}, 647
(2008)

\bibitem{Zyczkowski_U} K. \.{Z}yczkowski and M. Ku\'{s}, Random
unitary matrices, J. Phys. A: Math. Gen. \textbf{27}, 4235 (1994)

\end{thebibliography}
\end{document}